\documentclass[aps,a,preprint,superscriptaddress]{revtex4-1} 

\usepackage{graphicx} 
\usepackage{natbib}

\begin{document}

\preprint{}

\title{Ultrafast Voltage Sampling using Single-Electron Wavepackets}

\author  {N. Johnson}
\email[]{nathan.johnson@npl.co.uk}
\affiliation{National Physical Laboratory, Hampton Road, Teddington, Middlesex TW11 0LW, United Kingdom}
\affiliation{London Centre for Nanotechnology, and Department of Electronic and Electrical Engineering, University College London, Torrington Place, London, WC1E 7JE, United Kingdom}
\author{J. D. Fletcher}
\affiliation{National Physical Laboratory, Hampton Road, Teddington, Middlesex TW11 0LW, United Kingdom}
\author{D.A. Humphreys}
\affiliation{National Physical Laboratory, Hampton Road, Teddington, Middlesex TW11 0LW, United Kingdom}
\author {P. See}
\affiliation{National Physical Laboratory, Hampton Road, Teddington, Middlesex TW11 0LW, United Kingdom}
\author{J.P. Griffiths}
\affiliation {Cavendish Laboratory, University of Cambridge, J.J. Thomson Avenue, Cambridge CB3 0HE, United Kingdom}
\author{G.A.C. Jones}
\affiliation{Cavendish Laboratory, University of Cambridge, J.J. Thomson Avenue, Cambridge CB3 0HE, United Kingdom}
\author{I. Farrer}
\altaffiliation [Present address: ]{Department of Electronic and Electrical Engineering, University of Sheffield, Mappin Street, Sheffield S1 3JD, United Kingdom}
\affiliation {Cavendish Laboratory, University of Cambridge, J.J. Thomson Avenue, Cambridge CB3 0HE, United Kingdom}
\author{D.A. Ritchie}
\affiliation{Cavendish Laboratory, University of Cambridge, J.J. Thomson Avenue, Cambridge CB3 0HE, United Kingdom}
\author{M. Pepper}
\affiliation{London Centre for Nanotechnology, and Department of Electronic and Electrical Engineering, University College London, Torrington Place, London, WC1E 7JE, United Kingdom}
\author{T.J.B.M Janssen}
\affiliation{National Physical Laboratory, Hampton Road, Teddington, Middlesex TW11 0LW, United Kingdom}
\author{M. Kataoka}
\affiliation{National Physical Laboratory, Hampton Road, Teddington, Middlesex TW11 0LW, United Kingdom}

\begin{abstract}

We demonstrate an ultrafast voltage sampling technique using a stream of electron wavepackets. 
Electrons are emitted from a single-electron pump and travel through electron waveguides towards a detector potential barrier. 
Our electrons sample an instantaneous voltage on the gate upon arrival at the detector barrier.
Fast sampling is achieved by minimising the duration that the electrons interact with the barrier, which can be made as small as a few picoseconds. 
The value of the instantaneous voltage can be determined by varying the gate voltage to match the barrier height to the electron energy, which is used as a stable reference. 
The test waveform can be reconstructed by shifting the electron arrival time against it.
We argue that this method has scope to increase the bandwidth of voltage sampling to 100~GHz and beyond.

\end{abstract}
\date{\today}
\pacs{}

\maketitle 

There is much interest in producing faster electronic devices for high-performance computing and large-volume data communication.
High frequency signal analysis becomes important in the testing and design of these high speed applications.  
The bandwidth of commercial sampling oscilloscopes using sampling gates are approaching 100~GHz \cite{tektronix, *keysight}.
The limiting factor to their bandwidth is the parasitic capacitance in the components\cite{O'Dell}.
While it is in principle straightforward to generate trigger pulses at picosecond length for sampling gates, parasitic loss limits the sampling bandwidth. 

\begin{figure} [t!]
	\includegraphics [width=10cm]{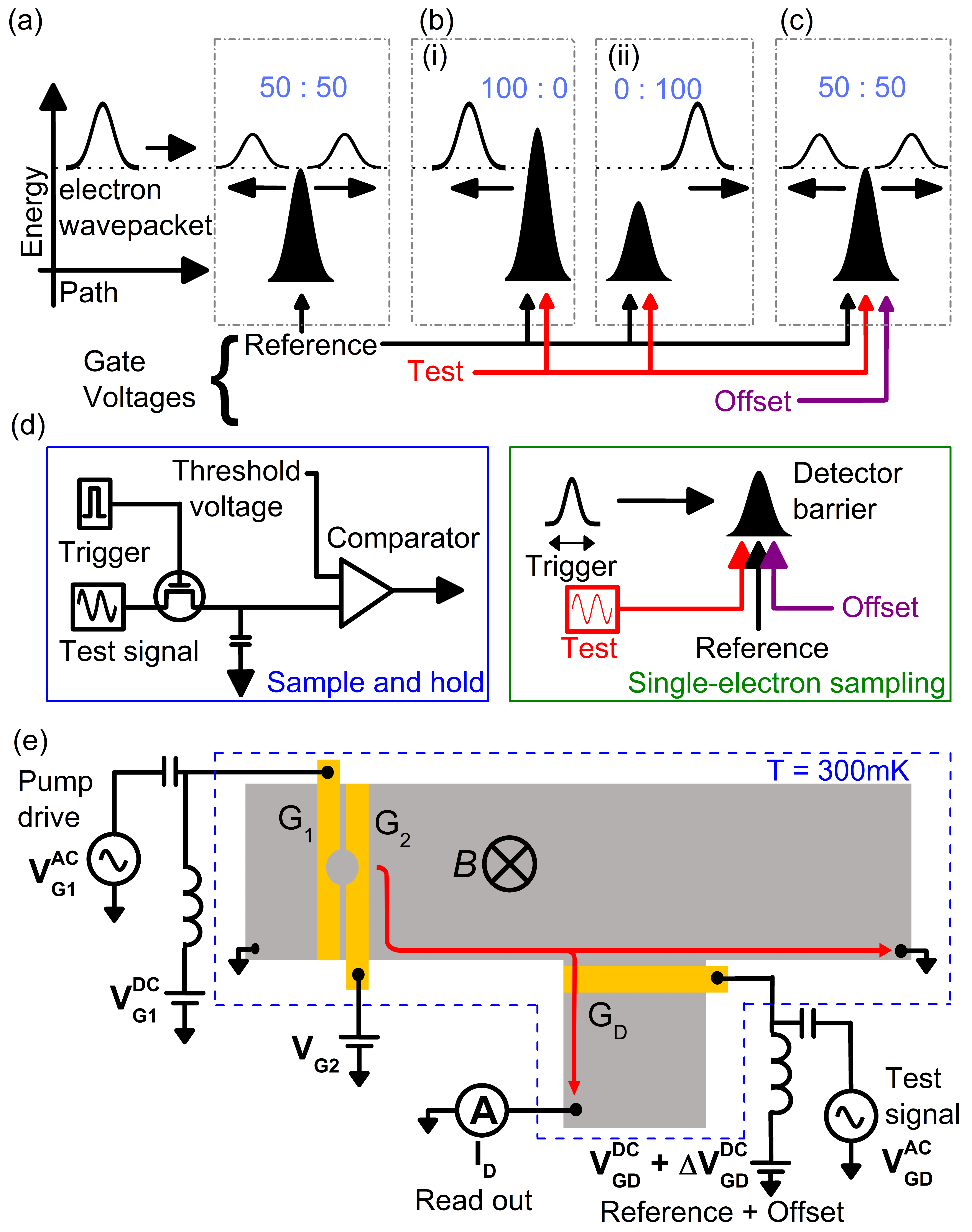}
	\label{fig1}
	\caption{
	\small
	(a)-(c) Principle of the single-electron-sampling (SES) scheme. 
	(a) We find the gate voltage (Reference) that gives half transmission through the barrier for incoming electron wavepackets.
	(b) When we add an unknown test signal transmission is modified to either (i) fully reflected or (ii) fully transmitted because the total potential on the gate is modified.
	(c) By adding the Offset voltage we return the detector gate potential to the half transmission point.
	(d) Block diagram comparisons between a sample-and-hold method and our SES method.
	(e) Schematic of the experimental setup. Gates G$_{\rm 1}$ and G$_{\rm 2}$ form the electron pump, which produces single-electron wavepackets. These propagate across the device  in edge states (red) to the detector barrier G$_{\rm D}$, which implements the SES scheme presented in (a)-(c).
	}
\end{figure}

There are various optical techniques developed to overcome the electrical bandwidth limitation \cite{ChouT, *FosterT}.
An alternative radical approach to such a problem may be to use a short wavepacket of a single quasiparticle (e.g. a conduction band electron) instead of a voltage trigger pulse.
The information transmitted by a quasiparticle wavefunction is protected in the absence of scattering or tunneling events. 
The use of this wavefunction as the media of (classical) information transfer would allow us to achieve high-speed device operations without the bandwidth limitations imposed by conventional transmission lines.  

In this letter, we present an ultrafast voltage sampling method using single-electron wavepackets traveling through electron waveguides in a semiconductor substrate.
An unknown test signal is added to a known gate voltage to form a potential barrier in the path of the electron wavepackets. 
The transmission probability through the barrier depends on the instantaneous barrier height on arrival at the detector relative to the electron energy. 
In this manner, our electrons can sample the test signal voltage, in a similar way to the sample and hold method using a voltage comparator in conventional sampling gates \cite{O'Dell}.  
High bandwidth of this sampling is achieved by tuning the electron's arrival-time distribution to 10 ps or shorter \cite{JoannaT}.  
Our method in principle eliminates the bandwidth limitation that plagues conventional electronic devices.
While we are presently limited by the bandwidth of the transmission line of the test signal, we argue that this method has the potential to increase the bandwidth of voltage sampling up to 100~GHz and beyond.

The principle of our single-electron-sampling (SES) scheme is presented in Figs.~1(a)-(c).
Single-electron wavepackets are generated at a fixed energy and travel along the same path towards a potential barrier, which we call the detector barrier.
The direction the wavepackets travel on arrival at the detector barrier depends on the barrier height, which is controlled by a gate voltage.
If the electrons have an energy greater than the potential on this barrier, they pass through it, otherwise they are deflected.
This path direction is only dependent on the instantaneous barrier potential at the time of electron arrival.
Therefore, we can use this information to sample the voltage at a very short timescale.

We initially apply a detector gate voltage such that the barrier is held at half transmission [Fig.~1(a)].
At this voltage, each electron has a $\sim$~50\% probability of tunneling through the detector potential.
We denote this threshold voltage the ``Reference".
When we add a test signal to the detector gate, the transmission of wavepackets either increases or decreases, depending on the sign of the test signal [Fig.~1(b)(i) and (ii)].
We then add a further dc voltage to the detector gate, which we call the ``Offset" [Fig.~1(c)]. 
If the magnitude of the Offset matches that of the instantaneous test signal, but with the opposite sign, then the barrier is brought back to the original point of half transmission. 
This way, we can ``sample'' the instantaneous value of the test signal.

The SES method can be compared to that of voltage sampling by a sample and hold method \cite{Pereira, O'Dell} as shown in the simplified block diagrams in Fig.~1(d).
In the sample and hold method, a trigger pulse closes a switch which permits the test signal to propagate onto a capacitor.
When the switch is opened, the voltage is fed into a comparator for analog-to-digital conversion.
This allows a fast waveform to be sampled by the capacitor taking ``snapshots" of the waveform each time it is charged.
In the SES method, single-electron wavepackets represent trigger pulses. 
The voltage comparison is made between the test signal and Offset voltage (with opposite signs), and the result is read by the direction that our electrons travel.
The waveform of the test signal can be scanned by shifting the electron arrival time against it to sample different parts of the waveform.
In principle, this method should work in single-shot mode (for real-time sampling), in which we could use a charge sensor to record the transmission of a single electron across the detector barrier.
Although technologically plausible, we do not yet have such capability.
Here, as a demonstration of proof of principle, we use a periodic test signal, to which the  timing of electron wavepacket emission is synchronised. 
Hence, the direction of electron flow can be detected as a current.

Fig.~1(e) shows a schematic of the device and connections used to realise the SES scheme.
A GaAs/AlGaAs heterostructure defines a two dimensional electron gas (2DEG) $\sim$90~nm below the surface.
The active area of the device is etched to confine the 2DEG to a 1.5~$\mu$m-wide channel (grey shading).
Ti/Au gates G$_{\rm 1}$, G$_{\rm 2}$ and G$_{\rm D}$ are patterned on the surface.
Gates  G$_{\rm 1}$ and  G$_{\rm 2}$ define the single-electron pump\cite{BlumenthalT, KaestnerT, GiblinT}, which is our source of electron wavepackets\cite{LeichtT, JonT, JoannaT, masaya2T}.
Gate  G$_{\rm D}$ forms the detector barrier 4~$\mu$m  away from the pump.
Experiments are performed in a cryostat with a base temperature $\sim$300~mK and with a perpendicular magnetic field of $B=14$~T
\footnote{We used such extreme experimental conditions (low temperature and high magnetic field) to maximise the electron scattering length, but this technique should in principle work at less extreme conditions ($\sim$1.5~K and $\sim$8~T should be sufficient)}.

The electron pump is operated so that it emits electrons one by one at a stable fixed energy $\sim$~100~meV above the Fermi energy, with typical broadening $\sim$~4~meV [the full width at half maximum (FWHM)] \cite{JonT}. 
G$_{\rm 1}$ is driven by an ac sinusoidal waveform $V_{G1}^{AC}$ at 240~MHz with peak-to-peak amplitude $\sim$~1~V from one channel of an arbitrary waveform generator (AWG) \footnote{Our pump drive and test waveforms are constructed with a 50 point sine wave at 12~GS/s, giving us a 240~MHz signal.}.
During the pump cycle, an electron populates the quantum dot, and is then ejected over the drain barrier G$_{\rm 2}$.
This produces a quantised current $I = ef$, with $e$ the elementary charge and $f$ the frequency of the driving waveform.
In the presence of a sufficiently large $B$, the ejected electrons travel along the sample edge [marked as red paths in Fig.~1(e)] as in the edge-state transport in the quantum Hall regime\cite{Halperin}, but as hot electrons in the states higher than the Fermi energy.
No appreciable energy loss occurs between the pump and the detector due to a long scattering length of order tens of microns \cite{TaubertT, JonT, CliveT, masaya2T}.

A test signal $V_{GD}^{AC}$ is applied to the detector from the second channel of the AWG, synchronised to the pump signal (for this work, we use a two-channel Tektronix AWG7122C, but in principle any synchronised RF source could be used).
The pump drive signal and, for this first test the detector test signal, are filtered using a 630~MHz low pass filter.
We place an ammeter on the ohmic contact behind the detector [see Fig.~1(e)] so that it records the detector transmission as the transmitted current $I_{D}$. 
We set the Reference, $V_{GD}^{DC}$ and Offset $\Delta V_{GD}^{DC}$ voltages on the detector gate as shown in Fig.~1(a)-(c).
We track the half transmission point by adjusting the Offset, and deduce the instantaneous voltage of the test waveform.

Fig.~2(a) shows how we sample the test waveform at different times to build up its temporal form.
Changing the delay between $V_{G1}^{AC}$ and $V_{GD}^{AC}$ by a quantity $\Delta t_{d}$ allows us to control the arrival time of the electrons at the detector.
Electrons will then sample a different part of the test waveform.
We can control $\Delta t_{d}$ with 1 ps resolution, using the internal skew control between the two output channels of the AWG. 
Because the electron arrival time distribution is so short compared to the timescale that our test waveform voltage changes, we consider the waveform to be quasi-static during the sampling time.   

Fig.~2(b) is an example result, with a filtered 240~MHz sine wave as the test waveform
\footnote{Although we use the same frequency for the pump drive and the test signal, the necessary condition for this scheme to work is only that the frequency of the test signal must be exactly an integer multiple of the pump drive frequency.
The electron pump has been shown to operate over a wide frequency range (typically $\sim$0.1~-~1~GHz), providing a wide range of measurable frequencies.
}
plotting in the colour scale the derivative $dI_{D}/d \Delta V_{GD}^{DC}$ of the measured current with respect to the Offset voltage.
The vertical axis is the Offset $\Delta V_{GD}^{DC}$, and the horizontal axis is $\Delta t_{d}$.
We perform a Gaussian peak fit to this derivative and plot its peak centre as a filled square in Fig.~2(c)
\footnote{We take the point of maximum derivative of the current with respect to the detector $dI_{D}/dV_{GD}^{DC}$ as our half transmission point. The point of maximum derivative may not occur at exactly half transmission, but this does not compromise our method. 
}.
Inverting the sign on the $\Delta V_{GD}^{DC}$-scale gives the measured waveform.
To examine the linearity of this method, we fit a sine curve (red line) to the experimental data points in Fig.~2(c). 
The residual of the fit is plotted in the inset to Fig.~2(c).
The standard deviation of the residual is 160~$\mu$V, and suggests that this method has a good linearity at this level (assuming that the AWG output and our dc gate voltage source, a Keithley 213, have good linearity).


\begin{figure} [t!]
	\includegraphics[width=10cm]{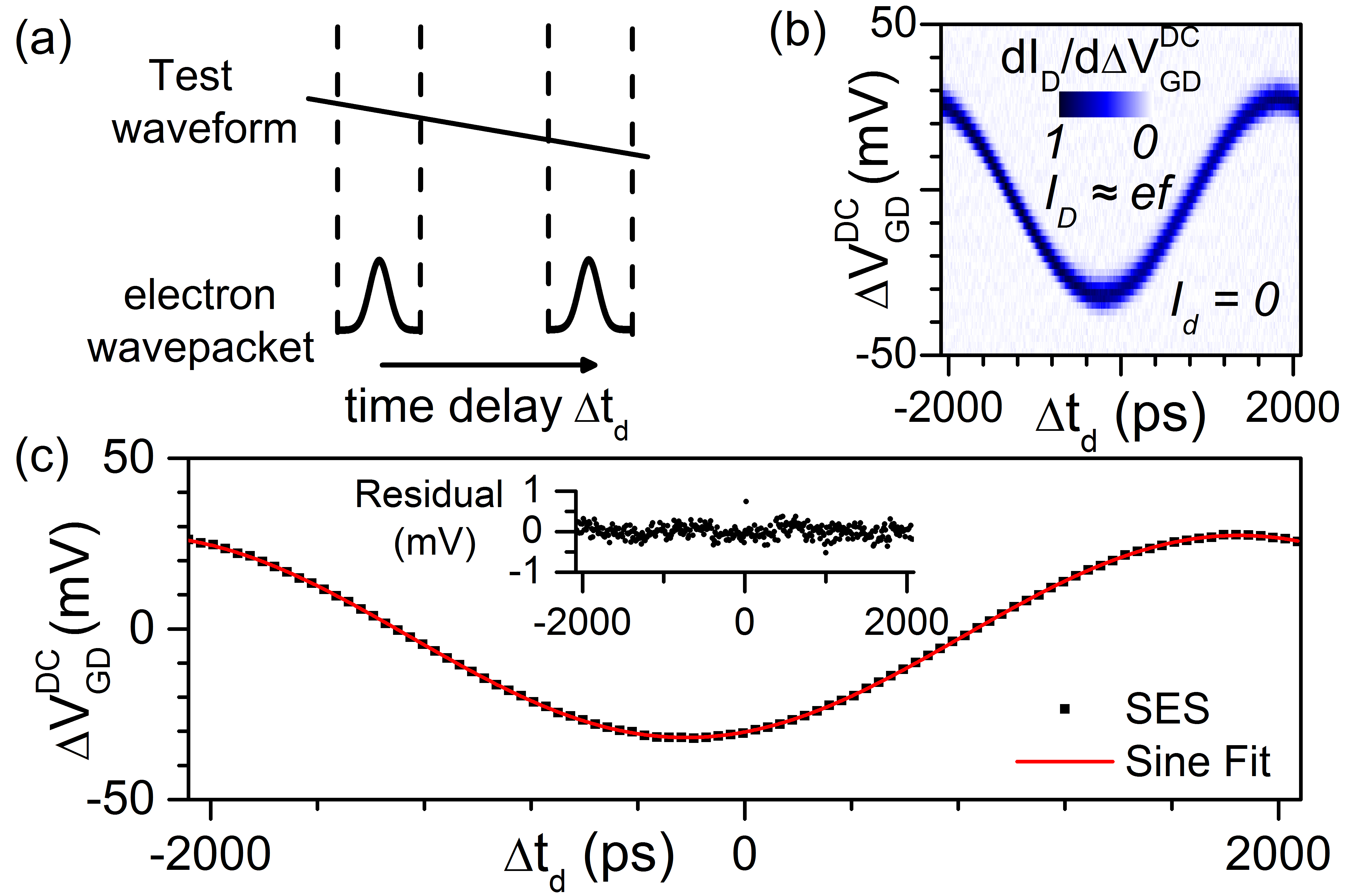}
	\caption{(a) Changing the time delay, $\Delta t_{d}$, allows the electron wavepacket to sample different parts of the test waveform. The sampling window is proportional to the length of the wavepacket in the time domain.
	(b) The derivative $dI_{D}/d\Delta V_{GD}^{DC}$ plotted against $\Delta V_{GD}^{DC}$ and $\Delta t_{d}$, which indicates the electron transmission threshold. The detector current $I_{D}= 0$ below this threshold (more negative $\Delta V_{GD}^{DC}$ while $I_{D} \approx ef$ above the threshold.
	(c) Filled squares show the peak centre of a Gaussian fit to the experimental data in (b). The red curve is a sine fit to the peak centre.
	Inset: The residual of the fit, implying good linearity at this voltage scale.
	}
\end{figure}	

\begin{figure} [t!]
	\includegraphics[width=10cm]{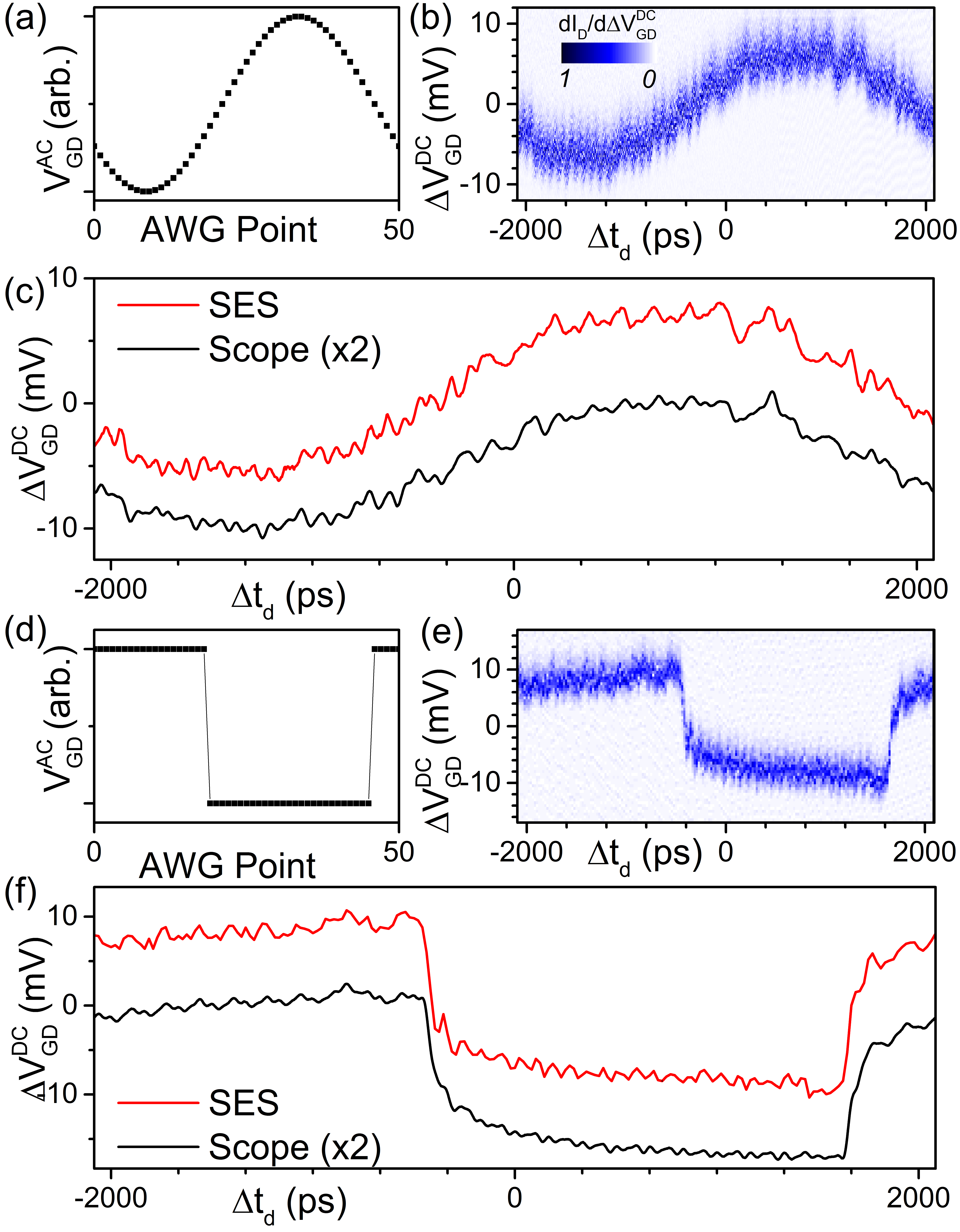}
	\caption{Measurements of unfiltered AWG outputs for (a)-(c) sine waveform and (d)-(f) square waveform. (a) and (d) show AWG waveform data. (b) and (e) show the detector derivative map as in Fig.~2(b). (c) and (f) show the waveform measured by the SES method (red) and the one measured by an oscilloscope (black). Note that for the oscilloscope traces, the sign of the voltage is inverted, the amplitude is doubled, and the trace is offset for comparison with the SES traces.}
\end{figure}

\begin{figure} [t!]
	\includegraphics[width=10cm]{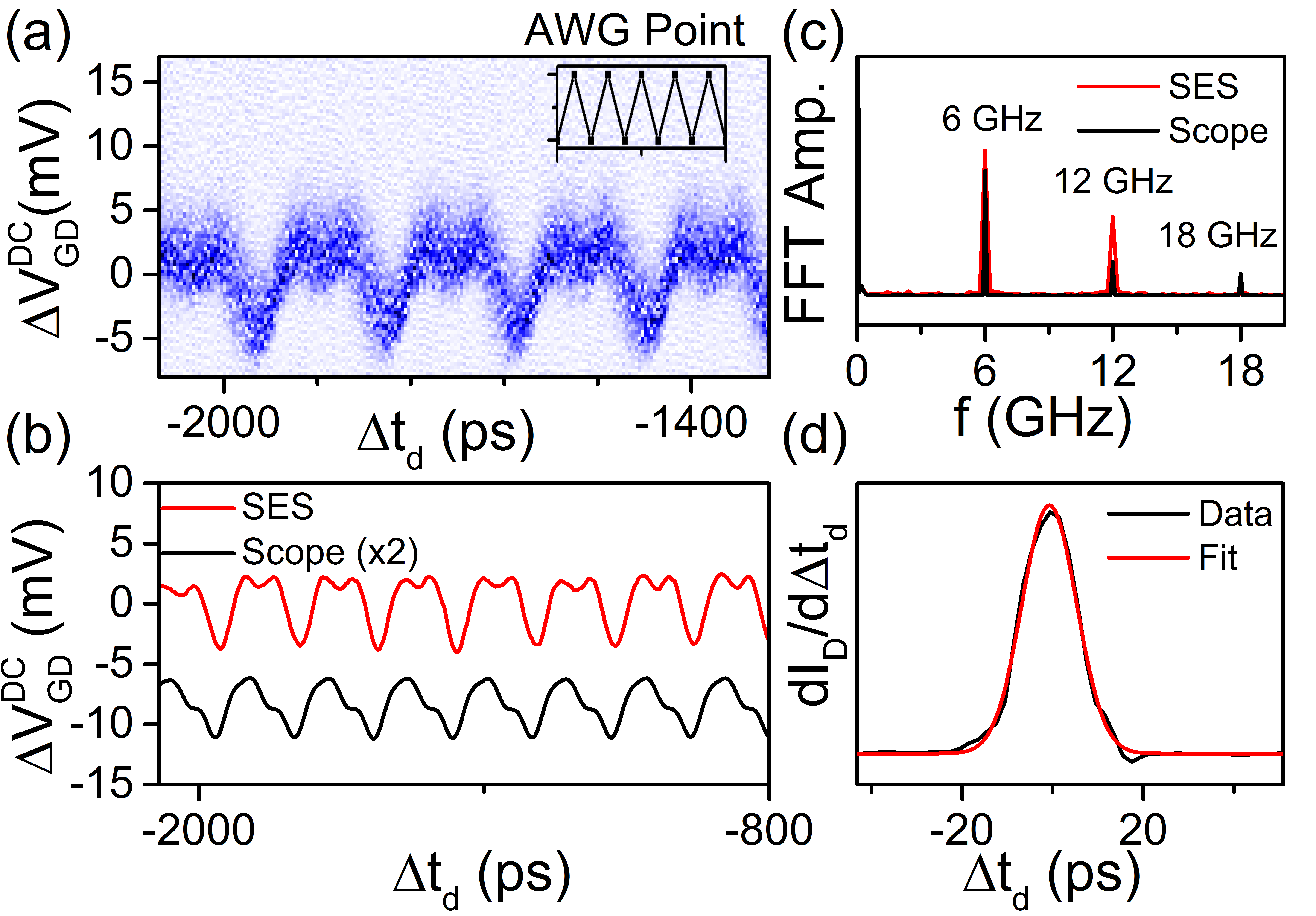}
	\caption{ 
	(a) Detector current derivative map measured for 6 GHz waveform constructed as shown in the inset. 
	(b) Comparison between the SES measurement (red) and the oscilloscope (black).
	(c) The Fast Fourier Transforms (FFT) of each trace taken from (b). The SES measurement records a larger 12~GHz peak but no 18~GHz peak in contrast to the oscilloscope.
	(d) Electron arrival-time distribution measurements following Ref.~\onlinecite{JoannaT}.  A Gaussian fit gives the FWHM to be 14 ps, giving a potential bandwidth of $\sim$35~GHz.
}
\end{figure}

The results mentioned above demonstrate the basic principle of the SES method.
We now explore its bandwidth limitation. 
We remove the filter from the test-signal line, so that the distortion by higher harmonics from the AWG transmit to the detector gate.
The AWG construction of a sine wave is shown in Fig.~3(a), and the resultant current map in Fig.~3(b), again plotted with the derivative of current in the colourscale for clarity, as in Fig.~2(b). 
Again, we extract the derivative-peak positions and plot them in Fig.~3(c) (red curve).
There are high-frequency distortions clearly visible, dominated by 12~GHz components at the AWG's sampling rate.

To compare these results against the conventional sampling method, we connect a Tektronix MSO72304DX Mixed Signal Oscilloscope (analog bandwidth 23~GHz) at the end of the measurement probe instead of our sample holder containing the device (at room temperature) \footnote{with an extra 1~m length of coaxial cable, but this does not affect the bandwidth significantly.}. We measure the AWG signal by the oscilloscope through the same signal line as the detector.
In Fig.~3(c), we compare the oscilloscope trace (black curve) with the SES result.
Because the oscilloscope measurement ($50~\Omega$ termination) has an amplitude of approximately half the magnitude of the SES scheme (open ended), we scale the scope trace by factor 2 to make it easier to compare the traces.
We also inverted the sign of the voltage  for the oscilloscope trace, as the SES trace is inverted when plotted in the Offset voltage.
While the overall features are similar, the SES trace show stronger higher harmonic signal.
In Figs.~3(d)-(f), we repeat the same analysis for a 240~MHz unfiltered square wave.
Again, the higher harmonic features are stronger for the SES trace.

In order to investigate high-frequency response of the SES system further, we use the two-point waveform construction to generate highest-frequency oscillations (6~GHz) as shown in the inset to Fig.~4(a) (the single-electron pump frequency is kept at 240~MHz). 
We obtain the current measurement map in Fig.~4(a), and plot the peak-position extraction (red curve) and oscilloscope trace (black curve) in Fig.~4(b).
They show slightly differing waveform shapes.
We take the Fast Fourier Transform (FFT) amplitude of both traces, and plot them in Fig.~4(c).
The SES trace records the 12~GHz component almost twice as high as that of the oscilloscope trace, while the latter records the 18~GHz component which is not seen on the former trace.

We estimate that the SES method should have a bandwidth well in excess of 18~GHz. The arrival-time distribution \cite{JoannaT} of the electrons measured at the detector barrier is 14~ps (FWHM) as shown in Fig.~4(d).
This should give a temporal resolution of 28~ps, corresponding to a bandwidth of 35~GHz.
We speculate that the lack of 18~GHz peak in the SES FFT in Fig.~4(c) may be due to the bandwidth limitation of the transmission line on our sample holder and GaAs chip.
The results in Fig.~3 and Fig.~4 implies that the SES method has a higher bandwidth at 12~GHz. 
However, the oscilloscope used should have a flat passband up to $\sim$11.5~GHz. 
Since it is unlikely that our bandwidth can be better, we speculate that the excess amplitude at 12~GHz may be due to an unidentified non-linear effect, although we do not see such an effect in the lower-frequency data [Fig.~2(c)].  
Further studies are needed to clarify the bandwidth performance of the SES method.
Recent work on the temporal wavepacket size indicates that with careful tuning of the pump operation conditions, it is plausible to generate wavepackets with arrival-time distribution less than 10~ps, or even 1~ps \cite{KataokaT}.
This opens the possibility of a bandwidth in excess of 100~GHz if the bandwidth of the detector transmission line can be improved.

To summarise, we have demonstrated a technique of using single-electron wavepackets to sample an unknown test waveform. 
This method is analogous to that employed in a sampling oscilloscope, but with the possibility of realising a bandwidth in excess of 100~GHz.
Other than high-bandwidth applications, one area which this method can be useful is in-situ voltage waveform measurements in a cryogenic environment. 
On-chip signal verification is becoming increasingly important for fine control of quantum systems, for example in qubit state initialisation \cite{HofheinzT}. 
Our system might be useful in such applications, as a way of verifying the shape of signals on chip, and opens up the possibility of quantum measurements through precise signal control.

We thank J.~Marrinan, C.~Skach, and J.~Niblock from Tektronix for useful discussions on AWG and sampling oscilloscope technologies. 
We also thank S.~P.~Giblin for useful discussions.  
This research was supported by the UK Department for Business, Energy, and Industrial Strategy and by the UK EPSRC.

\bibliography{voltagecomparator}

\begin{thebibliography}{23}%
\makeatletter
\providecommand \@ifxundefined [1]{%
 \@ifx{#1\undefined}
}%
\providecommand \@ifnum [1]{%
 \ifnum #1\expandafter \@firstoftwo
 \else \expandafter \@secondoftwo
 \fi
}%
\providecommand \@ifx [1]{%
 \ifx #1\expandafter \@firstoftwo
 \else \expandafter \@secondoftwo
 \fi
}%
\providecommand \natexlab [1]{#1}%
\providecommand \enquote  [1]{``#1''}%
\providecommand \bibnamefont  [1]{#1}%
\providecommand \bibfnamefont [1]{#1}%
\providecommand \citenamefont [1]{#1}%
\providecommand \href@noop [0]{\@secondoftwo}%
\providecommand \href [0]{\begingroup \@sanitize@url \@href}%
\providecommand \@href[1]{\@@startlink{#1}\@@href}%
\providecommand \@@href[1]{\endgroup#1\@@endlink}%
\providecommand \@sanitize@url [0]{\catcode `\\12\catcode `\$12\catcode
  `\&12\catcode `\#12\catcode `\^12\catcode `\_12\catcode `\%12\relax}%
\providecommand \@@startlink[1]{}%
\providecommand \@@endlink[0]{}%
\providecommand \url  [0]{\begingroup\@sanitize@url \@url }%
\providecommand \@url [1]{\endgroup\@href {#1}{\urlprefix }}%
\providecommand \urlprefix  [0]{URL }%
\providecommand \Eprint [0]{\href }%
\providecommand \doibase [0]{http://dx.doi.org/}%
\providecommand \selectlanguage [0]{\@gobble}%
\providecommand \bibinfo  [0]{\@secondoftwo}%
\providecommand \bibfield  [0]{\@secondoftwo}%
\providecommand \translation [1]{[#1]}%
\providecommand \BibitemOpen [0]{}%
\providecommand \bibitemStop [0]{}%
\providecommand \bibitemNoStop [0]{.\EOS\space}%
\providecommand \EOS [0]{\spacefactor3000\relax}%
\providecommand \BibitemShut  [1]{\csname bibitem#1\endcsname}%
\let\auto@bib@innerbib\@empty
\bibitem [{\citenamefont {Tektronix}()}]{tektronix}%
  \BibitemOpen
  \bibfield  {author} {\bibinfo {author} {\bibnamefont {Tektronix}},\
  }\href@noop {} {\enquote {\bibinfo {title} {Techniques for extending
  real-time oscilloscope bandwidth},}\ }\BibitemShut {NoStop}%
\bibitem [{\citenamefont {{Keysight Technologies}}(2013)}]{keysight}%
  \BibitemOpen
  \bibfield  {author} {\bibinfo {author} {\bibnamefont {{Keysight
  Technologies}}},\ }\href
  {http://literature.cdn.keysight.com/litweb/pdf/5989-8794EN.pdf?id=1464210}
  {\enquote {\bibinfo {title} {What is the difference between an equivalent
  time sampling oscilloscope and a real-time oscilloscope? (application
  note)},}\ } (\bibinfo {year} {2013})\BibitemShut {NoStop}%
\bibitem [{\citenamefont {O'Dell}(1991)}]{O'Dell}%
  \BibitemOpen
  \bibfield  {author} {\bibinfo {author} {\bibfnamefont {T.~H.}\ \bibnamefont
  {O'Dell}},\ }\href@noop {} {\emph {\bibinfo {title} {Circuits for Electronic
  Instrumentation}}}\ (\bibinfo  {publisher} {Cambridge University Press},\
  \bibinfo {year} {1991})\BibitemShut {NoStop}%
\bibitem [{\citenamefont {Chou}\ \emph {et~al.}(2007)\citenamefont {Chou} \emph
  {et~al.}}]{ChouT}%
  \BibitemOpen
  \bibfield  {author} {\bibinfo {author} {\bibfnamefont {J.}~\bibnamefont
  {Chou}} \emph {et~al.},\ }\href@noop {} {\bibfield  {journal} {\bibinfo
  {journal} {App. Phys. Lett. 91, 161105}\ } (\bibinfo {year}
  {2007})}\BibitemShut {NoStop}%
\bibitem [{\citenamefont {Foster}\ \emph {et~al.}(2008)\citenamefont {Foster}
  \emph {et~al.}}]{FosterT}%
  \BibitemOpen
  \bibfield  {author} {\bibinfo {author} {\bibfnamefont {M.~A.}\ \bibnamefont
  {Foster}} \emph {et~al.},\ }\href@noop {} {\bibfield  {journal} {\bibinfo
  {journal} {Nature 456 81}\ } (\bibinfo {year} {2008})}\BibitemShut {NoStop}%
\bibitem [{\citenamefont {Waldie}\ \emph {et~al.}(2015)\citenamefont {Waldie}
  \emph {et~al.}}]{JoannaT}%
  \BibitemOpen
  \bibfield  {author} {\bibinfo {author} {\bibfnamefont {J.}~\bibnamefont
  {Waldie}} \emph {et~al.},\ }\href@noop {} {\bibfield  {journal} {\bibinfo
  {journal} {Phys. Rev. B 92, 125305}\ } (\bibinfo {year} {2015})}\BibitemShut
  {NoStop}%
\bibitem [{\citenamefont {Pereira}(2006)}]{Pereira}%
  \BibitemOpen
  \bibfield  {author} {\bibinfo {author} {\bibfnamefont {J.~M.~D.}\
  \bibnamefont {Pereira}},\ }\href@noop {} {\bibfield  {journal} {\bibinfo
  {journal} {IEEE Instr. Meas. Mag. 9 27}\ } (\bibinfo {year}
  {2006})}\BibitemShut {NoStop}%
\bibitem [{\citenamefont {Blumenthal}\ \emph {et~al.}(2007)\citenamefont
  {Blumenthal} \emph {et~al.}}]{BlumenthalT}%
  \BibitemOpen
  \bibfield  {author} {\bibinfo {author} {\bibfnamefont {M.~D.}\ \bibnamefont
  {Blumenthal}} \emph {et~al.},\ }\href@noop {} {\bibfield  {journal} {\bibinfo
   {journal} {Nature Phys. 3 343}\ } (\bibinfo {year} {2007})}\BibitemShut
  {NoStop}%
\bibitem [{\citenamefont {Kaestner}\ \emph {et~al.}(2008)\citenamefont
  {Kaestner} \emph {et~al.}}]{KaestnerT}%
  \BibitemOpen
  \bibfield  {author} {\bibinfo {author} {\bibfnamefont {B.}~\bibnamefont
  {Kaestner}} \emph {et~al.},\ }\href@noop {} {\bibfield  {journal} {\bibinfo
  {journal} {App. Phys. Lett. 92 192106}\ } (\bibinfo {year}
  {2008})}\BibitemShut {NoStop}%
\bibitem [{\citenamefont {Giblin}\ \emph {et~al.}(2012)\citenamefont {Giblin}
  \emph {et~al.}}]{GiblinT}%
  \BibitemOpen
  \bibfield  {author} {\bibinfo {author} {\bibfnamefont {S.~P.}\ \bibnamefont
  {Giblin}} \emph {et~al.},\ }\href@noop {} {\bibfield  {journal} {\bibinfo
  {journal} {Nature Comms. 1935}\ } (\bibinfo {year} {2012})}\BibitemShut
  {NoStop}%
\bibitem [{\citenamefont {Leicht}\ \emph {et~al.}(2011)\citenamefont {Leicht}
  \emph {et~al.}}]{LeichtT}%
  \BibitemOpen
  \bibfield  {author} {\bibinfo {author} {\bibfnamefont {C.}~\bibnamefont
  {Leicht}} \emph {et~al.},\ }\href@noop {} {\bibfield  {journal} {\bibinfo
  {journal} {Semicond. Sci. Tech. 26 055010}\ } (\bibinfo {year}
  {2011})}\BibitemShut {NoStop}%
\bibitem [{\citenamefont {Fletcher}\ \emph {et~al.}(2013)\citenamefont
  {Fletcher} \emph {et~al.}}]{JonT}%
  \BibitemOpen
  \bibfield  {author} {\bibinfo {author} {\bibfnamefont {J.~D.}\ \bibnamefont
  {Fletcher}} \emph {et~al.},\ }\href@noop {} {\bibfield  {journal} {\bibinfo
  {journal} {Phys. Rev. Lett. 111 216807}\ } (\bibinfo {year}
  {2013})}\BibitemShut {NoStop}%
\bibitem [{\citenamefont {Kataoka}\ \emph {et~al.}(2016)\citenamefont {Kataoka}
  \emph {et~al.}}]{masaya2T}%
  \BibitemOpen
  \bibfield  {author} {\bibinfo {author} {\bibfnamefont {M.}~\bibnamefont
  {Kataoka}} \emph {et~al.},\ }\href@noop {} {\bibfield  {journal} {\bibinfo
  {journal} {Phys. Rev. Lett. 116, 126803}\ } (\bibinfo {year}
  {2016})}\BibitemShut {NoStop}%
\bibitem [{Note1()}]{Note1}%
  \BibitemOpen
  \bibinfo {note} {We used such extreme experimental conditions (low
  temperature and high magnetic field) to maximise the electron scattering
  length, but this technique should in principle work at less extreme
  conditions ($\sim $1.5~K and $\sim $8~T should be sufficient)}\BibitemShut
  {NoStop}%
\bibitem [{Note2()}]{Note2}%
  \BibitemOpen
  \bibinfo {note} {Our pump drive and test waveforms are constructed with a 50
  point sine wave at 12~GS/s, giving us a 240~MHz signal.}\BibitemShut {Stop}%
\bibitem [{\citenamefont {Halperin}(1982)}]{Halperin}%
  \BibitemOpen
  \bibfield  {author} {\bibinfo {author} {\bibfnamefont {B.~I.}\ \bibnamefont
  {Halperin}},\ }\href@noop {} {\bibfield  {journal} {\bibinfo  {journal}
  {Phys. Rev. B 25 2185}\ } (\bibinfo {year} {1982})}\BibitemShut {NoStop}%
\bibitem [{\citenamefont {Taubert}\ \emph {et~al.}(2011)\citenamefont {Taubert}
  \emph {et~al.}}]{TaubertT}%
  \BibitemOpen
  \bibfield  {author} {\bibinfo {author} {\bibfnamefont {D.}~\bibnamefont
  {Taubert}} \emph {et~al.},\ }\href@noop {} {\bibfield  {journal} {\bibinfo
  {journal} {Phys. Rev. B 83, 235404}\ } (\bibinfo {year} {2011})}\BibitemShut
  {NoStop}%
\bibitem [{\citenamefont {Emary}\ \emph {et~al.}(2016)\citenamefont {Emary}
  \emph {et~al.}}]{CliveT}%
  \BibitemOpen
  \bibfield  {author} {\bibinfo {author} {\bibfnamefont {C.}~\bibnamefont
  {Emary}} \emph {et~al.},\ }\href@noop {} {\bibfield  {journal} {\bibinfo
  {journal} {Phys. Rev. B 93, 035436}\ } (\bibinfo {year} {2016})}\BibitemShut
  {NoStop}%
\bibitem [{Note3()}]{Note3}%
  \BibitemOpen
  \bibinfo {note} {Although we use the same frequency for the pump drive and
  the test signal, the necessary condition for this scheme to work is only that
  the frequency of the test signal must be exactly an integer multiple of the
  pump drive frequency. The electron pump has been shown to operate over a wide
  frequency range (typically $\sim $0.1~-~1~GHz), providing a wide range of
  measurable frequencies.}\BibitemShut {Stop}%
\bibitem [{Note4()}]{Note4}%
  \BibitemOpen
  \bibinfo {note} {We take the point of maximum derivative of the current with
  respect to the detector $dI_{D}/dV_{GD}^{DC}$ as our half transmission point.
  The point of maximum derivative may not occur at exactly half transmission,
  but this does not compromise our method.}\BibitemShut {Stop}%
\bibitem [{Note5()}]{Note5}%
  \BibitemOpen
  \bibinfo {note} {With an extra 1~m length of coaxial cable, but this does not
  affect the bandwidth significantly.}\BibitemShut {Stop}%
\bibitem [{\citenamefont {Kataoka}\ \emph {et~al.}()\citenamefont {Kataoka}
  \emph {et~al.}}]{KataokaT}%
  \BibitemOpen
  \bibfield  {author} {\bibinfo {author} {\bibfnamefont {M.}~\bibnamefont
  {Kataoka}} \emph {et~al.},\ }\href@noop {} {\bibinfo  {journal} {to be
  published}\ }\BibitemShut {NoStop}%
\bibitem [{\citenamefont {Hofheinz}\ \emph {et~al.}(2009)\citenamefont
  {Hofheinz} \emph {et~al.}}]{HofheinzT}%
  \BibitemOpen
\bibfield  {journal} {  }\bibfield  {author} {\bibinfo {author} {\bibfnamefont
  {M.}~\bibnamefont {Hofheinz}} \emph {et~al.},\ }\href@noop {} {\bibfield
  {journal} {\bibinfo  {journal} {Nature 459 548}\ } (\bibinfo {year}
  {2009})}\BibitemShut {NoStop}%
\end{thebibliography}%

\end{document}